\documentclass{article}
\usepackage{spconf,amsmath,graphicx}

\usepackage{hyperref}
\usepackage{caption}

\usepackage{amssymb}
\usepackage{multirow} 
\graphicspath{{grf/}}

\newcommand{\V}{\ensuremath{\mathbf{V}}}
\newcommand{\W}{\ensuremath{\mathbf{W}}}
\newcommand{\X}{\ensuremath{\mathbf{X}}}

\newcommand{\h}{\ensuremath{\mathbf{h}}}

\newcommand{\sss}{\ensuremath{\mathbf{s}}}  

\newcommand{\vv}{\ensuremath{\mathbf{v}}}

\newcommand{\x}{\ensuremath{\mathbf{x}}}

\newcommand{\bbR}{\ensuremath{\mathbb{R}}}

\newcommand{\caja}[4][1]{{%
    \renewcommand{\arraystretch}{#1}%
    \begin{tabular}[#2]{@{}#3@{}}%
      #4%
    \end{tabular}%
    }}

{%
\begin{list}{#1}{
\vspace{-\topsep}
\vspace{-\partopsep}
\setlength{\itemindent}{0cm}
\setlength{\rightmargin}{0cm}
\setlength{\listparindent}{0cm}
\settowidth{\labelwidth}{#1}
\setlength{\leftmargin}{\labelwidth}
\addtolength{\leftmargin}{\labelsep}
\setlength{\itemsep}{0cm}
}%
}%
{%
\end{list}
\vspace{-\topsep}
\vspace{-\partopsep}
}

{\begin{enumerate}%
}%
{\end{enumerate}}

\newcommand{\sbr}[1]{\left[#1\right]}
\newcommand{\rbr}[1]{\left(#1\right)}
\newcommand{\cbr}[1]{\left\{#1\right\}}

\DeclareMathOperator*{\argmax}{arg\,max}

\usepackage{color}
\usepackage{colortbl}
\definecolor{light-gray}{gray}{0.8}

\title{Acoustic Scene Analysis with Multi-head Attention Networks}

\name{Weimin Wang, Weiran Wang,  Ming Sun,  Chao Wang}
\address{Amazon Alexa  \\ 101 Main St, Cambridge, MA 02142, USA \\
\texttt{\{wanweimi,weiranw,mingsun,wngcha\}@amazon.com} }

\begin{document}
\maketitle

\begin{abstract}
Acoustic Scene Classification (ASC) is a challenging task, as a single scene may involve multiple events that contain complex sound patterns. For example, a cooking scene may contain several sound sources including silverware clinking, chopping, frying, etc. What complicates ASC more is that classes of different activities could have overlapping sounds patterns (e.g. both cooking and dishwashing could have silverware clinking sound). In this paper, we propose a multi-head attention network to model the complex temporal input structures for ASC. The proposed network takes the audio's time-frequency representation as input, and it leverages standard VGG plus LSTM layers to extract high-level feature representation. Further more, it applies multiple attention heads to summarize various patterns of sound events into fixed dimensional representation, for the purpose of final scene classification. The whole network is trained in an end-to-end fashion with backpropagation. Experimental results confirm that our model discovers meaningful sound patterns through the attention mechanism, without using explicit supervision in the alignment. We evaluated our proposed model using DCASE 2018 Task 5 dataset, and achieved competitive performance on par with previous winner's results.  
\end{abstract}


\section{Introduction}
\label{s:intro}


High level semantic understanding of an audio stream is a fundamental problem in machine intelligence. Being able to infer from sound patterns what events are occuring and what is the surrounding environment has potential applications in a wide range of fields such as public safety~\cite{ntalampiras2009acoustic}, ecological study~\cite{lammers2008ecological}, and assisted living~\cite{schroeder2011detection}.

Recently, the tasks of audio event detection (AED) and acoustic scene analysis (ASC) have gained increasing popularity, due to the availability of large scale datasets~\cite{Gemmeke2017} and commonly used benchmarks~\cite{DCASE2017, DCASE2018}. 
We observe that, deep learning architectures such as convolutional neural networks~\cite{krizhevsky2012imagenet,He_2016} and long short-term memory networks~\cite{HochreitSchmid97a} and their variants have contributed significantly to the success of many approaches to the tasks. 

Although the common deep learning architectures can work well for fully supervised tasks, challenges arise when the task at hand is only weakly labeled, which is often the case in practice. As an example, a \emph{cooking} scene may contain several sound events including silverware clanking, chopping, frying, and perhaps other human activities (such as walking and talking); knowing the existence of the smaller events (with short temporal duration) clearly helps inferring the abstract scene class. To train a ASC system, it is challeging to collect datasets with fine-grained event labels: human annotators may quickly categorize the whole scene correctly, but it would be laborious (and also expensive) for them to exhaustively identify the smaller sound events and to pinpoint their onset/offset times. A more practical approach is to collect sufficient amount of recordings with only scene class, and develop models which exploit the structure that a scene typically consists of multiple smaller events, and perform recording-level inference based on the aggregation of evidence at the event level, with supervision only at the scene level. Due to both the challenges and opportunities, weakly supervised learning has been a continual scheme in this research area.

In this paper, we propose a multi-head attention network for ASC, which implements the abovementioned intuitions of hierarchical representation and compositional inference. Our model applies multiple attention heads to frame level representations of the input recording,  where each head has a hidden event in mind, and attends to relevent frames to extract a recording-level features; the features from all attention heads are then pooled together, as the final representation of the entire recording for scene classification. Although we only receive supervision at the recording level, experimental results show that our model discovers meaningful sound patterns through the attention mechanism, and the soft alignment provided by the attention heads encode high-quality time information. In the rest of this paper, we give detailed formulation of our method in Sec.~\ref{s:model}, discuss related work in Sec.~\ref{s:related}, present experimental results in Sec.~\ref{s:expt} and qualitative analysis of the attentions in Sec.~\ref{s:visualizations}, and conclude in Sec.~\ref{s:conclusion}.

\section{Multi-head attention for acoustic scene analysis}
\label{s:model}

Our task is to associate each audio clip with a scene class. The scene can be a
high level, abstract concept which consist of various smaller events. For
example, in a typical ``cooking'' scene, we expect to hear events like
cook-ware, cutting, dishwashing, and human activity sounds like walking and
talking. In another scene ``working'', we could hear events such as
keyboard typing and mouse clicks, as well as paper scratching. 

In order to categorize the overall scene, it can be helpful to
detect the existence of such smaller events and to analyze their
co-occurrence.
However, it is very costly to pre-define the set of smaller events and
have human annotate their occurrence in audio clips. In this section, we
propose a method for automatically detecting the existence of meaningful
events and locating their appearances in time (alignment) for scene classification.

Let an input utterance be $\X = [\x_1, \dots, \x_{T^\prime}]$ where $\x_i \in
\bbR^d$ contains the features for the $i$th audio frame, and $T^\prime$ is the
total number of frames. We apply
deep convolutional networks followed by bi-directional LSTMs to extract
high-level features that contain rich context information from the input
(see Section~\ref{s:architecture} for details).
Let the output of this feature extraction networks (denoted by $f$) be
$f(\X) = [\h_1, \dots, \h_{T}]$ where $\h_t \in \bbR^p$ and
$T\le T^\prime$ due to subsampling in the time axis.

For each input sequence, we consider a set of $M$ smaller events,
where $M$ is a hyper-parameter to be tuned by cross-validation.
Let the vectorial representation of the $i$th event be $\vv^i \in
\bbR^p$, 
and write representations of all events collectively as $\V =
\cbr{\vv^i}_{i=1,\dots,M}$.
We compute the similarity between the $\h_t$ sequence and
$\vv^i$, followed by exponentiation and normalization, to obtain the
``attention scores'' for event $i$:
\begin{align*}
  a^i_t = \exp ( \h_t^\top \vv^i / \sigma) / \sum_{t^\prime=1}^T {\exp (
  \h_{t^\prime}^\top \vv^i / \sigma)},
  \quad\text{for}\; t=1,\dots,T,
  \end{align*}
where $\sigma>0$ is a hyper-parameter that controls the sharpness of the
soft alignment (the smaller the $\sigma$ is, the more peaked the attention
scores are).
The nonnegative attention scores $\cbr{a^i_t}_{t=1,\dots,T}$ satisfy $\sum_{t=1}^T a^i_t=1$,
and highlight the most relevant frames (for event $i$) from $f(X)$,
while pushing the affinity of others frames close to zero.
We then summarize the feature sequence into a fixed dimensional vector
\begin{align*}
  \sss^i = \sum_{t=1}^T a^i_t \h_t \in \bbR^p, \quad i=1,\dots,M
\end{align*}
for each event. 
Finally, we concatenate all $M$ events' representations to obtain
\begin{align*}
  \sss = [\sss^1;\dots;\sss^M] \in \bbR^{Mp} ,
\end{align*}
and use it as the final feature for the entire utterance. 

For $N$-class scene classification, we apply a feed-forward network $g$ with a
final softmax layer to the utterance representation $\sss$.  with weights
$\W \in \bbR^{N\times Mp}$ at the end, to get predictions
\begin{align*}
  \sbr{P(y=1|\X), \dots ,P(y=N|\X)} = \text{softmax} \rbr{ g (\sss) }.
  \end{align*}
Given a training set of $(\X,y)$ pairs, we jointly learn parameters in
feature extraction network $f$, event representations
  $\V$, and classification network $g$ using the cross-entropy loss.

The attention mechanism is widely used in speech
recognition~\cite{Chorowski_15a,Chan_16a} 
and natural language processing~\cite{Bahdanau_15a}, and multi-head self-attention has
been proposed in~\cite{Vaswani_17a}.
In this work, we have borrowed the same intuition of learn-able, soft
alignment from these prior work for detecting events in an unsupervised
fashion, and our use of multiple attention heads is motivated by the complex
nature of scenes---each scene may contain several distinctive events. 
In related settings, a few recent work~\cite{Wang_18b,McFee_18a,Wang_19a} formulated the weakly supervised
event detection problem (given only the utterance label, train a system to
infer both utterance label and time alignment) as a multiple instance
learning problem, and proposed different pooling strategies to aggregate the
per-instance (or per-segment) hypothesis to form an utterance level
prediction, on top of which supervision is imposed. This aggregation process resembles attention, albeit at the
prediction score level for \emph{specific} target event, rather than at the representation level. 
A similar attention mechanism to ours was used for rare event
detection in~\cite{Wang_18a}, where the attention scores receives supervision from the
onset/offset time provided by their task (Challenge 2 of DCASE
2017~\cite{Mesaro_17a}).
In contrast, we do not have any supervision in this work for the
frame-wise alignment. 
Instead, we rely on the model's structural constraint---the multi-head attention
mechanism---to attend to multiple relevant snippets and combine them for
classification.
To encourage the model to discover diverse events, we apply
dropout~\cite{Srivas_14a} throughout the model (and in particular on $\sss$) to prevent the
$\vv^i$'s from co-adaptation. As we will see in the empirical analysis,
our model automatically discovers events types that are semantically meaningful, and are highly correlated to the scene
classes. 

\section{Related work}
\label{s:related}


Here we briefly describe a few previous approaches on DCASE 2018 Task 5, which we will use in the experiments. The baseline system provided by the organizer~\cite{Kong2018} was based on a 1-D CNN model applied to input Mel-spectrogram features extracted from the original 10-second clips. They treated each of the 4 channels of the audio clip as one independent data point during training, and this strategy was adopted by most of the teams. 
Among the top winning teams of the challenge,~\cite{Inoue2018} similarly adopted a CNN-based architecture, except that they applied 2D convolutions along both the time and frequency dimension, which gave significant improvement over the baseline results. 
~\cite{tanabe2018multichannel} applied heavy feature engineering and pre-processing techniques, such as blind dereverberation, blind source separation, and noise reduction, as their ‘Front-End Modules’, which can potentially be useful to our model as well.~\cite{Liao2018} applied sub-band convolutions in their architecture, and performed frame-wise prediction of scene label, with frame targets generated heuristically based on energy.~\cite{Nakadai2018} learned one shared network from two task---scene classification and regression, where the regression task is to predict pre-computed single-channel representations from multi-channel input data, and showed that the regression task helps improve classification performance. 
\section{Experimental results}
\label{s:expt}

\begin{table}[t]
\centering
\begin{tabular}{@{}|l|r|r|@{}}
\hline
Activity & \#10s seg & \# sessions \\
\hline
Absence (nobody in room) & 18860 & 42 \\
Cooking & 5124 & 13 \\
Dishwashing & 1424 & 10 \\
Eating & 2308 & 13 \\
Other (no relevant activity) & 2060 & 118 \\
Social activity (visit, phone call) & 4944 & 21 \\
Vacuum cleaning & 972 & 9 \\
Watching TV & 18648 & 9 \\
Working (typing, mouse click, ...) & 18644 & 33 \\
\hline
Total & 72984 & 268 \\
\hline
\end{tabular}
\caption{Class distributions with number of training samples and recording
  sessions.}
\label{t:class-distribution}
\end{table}

\subsection{Dataset}

We demonstrate the proposed method on the task 5 of DCASE 2018 challenge~\cite{Dekkers2018_DCASE}, an acoustic
scene classification task derived from the SINS
dataset~\cite{dekkers2017sins}. 
In this dataset, each input audio clip is 10 seconds long, consisting of
four acoustic channels with the sampling rate of 16 kHz. 

We ignore the correlation between different channels, and treat the data
from each channel as independent for training, as is done in previous work~\cite{tanabe2018multichannel};
this yields a 4x augmentation of the training set size from 73K to
292K. 
During inference, for each audio clip, we obtain the predictions for each channel, and
average the four prediction scores for the final classification.

\begin{table}[t]
  \caption{Configuration of the feature extraction network in proposed model.
    Note that all Conv layers below contain batch normalization~\cite{Ioffe_15a} and ReLU activation.}
  \label{t:convnet}
  \begin{tabular}{@{}|l|c|c|r|r|@{}}
    \hline
    layer & 	         kernel size &       	stride        &    \# filters &	      data shape \\
    \hline
    Input         &                           &                             &                     &             (64, 1250) \\
    \hline
    Conv         &           3x3           &               1x1        &               64  & \\ 
    Conv         &           3x3           &               1x1        &               64  & \\
    MaxPool    &           2x2           &               2x2        &                     &           (32, 625, 64) \\
    \hline
    Conv         &           3x3           &               1x1        &              128 & \\
    Conv         &           3x3           &               1x1        &              128 & \\
    MaxPool    &           2x2           &               2x2        &                     &           (16, 312, 128) \\
    \hline
    Conv         &           3x3           &               1x1        &              256 & \\
    Conv         &           3x3           &               1x1        &              256 & \\
    Conv         &           3x3           &               1x1        &              256 & \\
    MaxPool    &           2x2           &               2x2        &                     &            (8, 156, 256) \\
    \hline
    Conv         &           3x3           &               1x1        &              512 & \\
    Conv         &           3x3           &               1x1        &              512 & \\
    Conv         &           3x3           &               1x1        &              512 & \\
    MaxPool    &           2x1           &               2x1        &                     &            (4, 156, 512) \\
    \hline
    Conv         &           3x3           &               1x1        &              512 & \\
    Conv         &           3x3           &               1x1        &              512 & \\
    Conv         &           3x3           &               1x1        &              512 & \\
    MaxPool    &           2x1           &               2x1        &                     &            (2, 156, 512) \\
    \hline
  \end{tabular}
\end{table}

\begin{figure}[t]
  \centering
  \includegraphics[width=0.75\linewidth,height=1.2\linewidth]{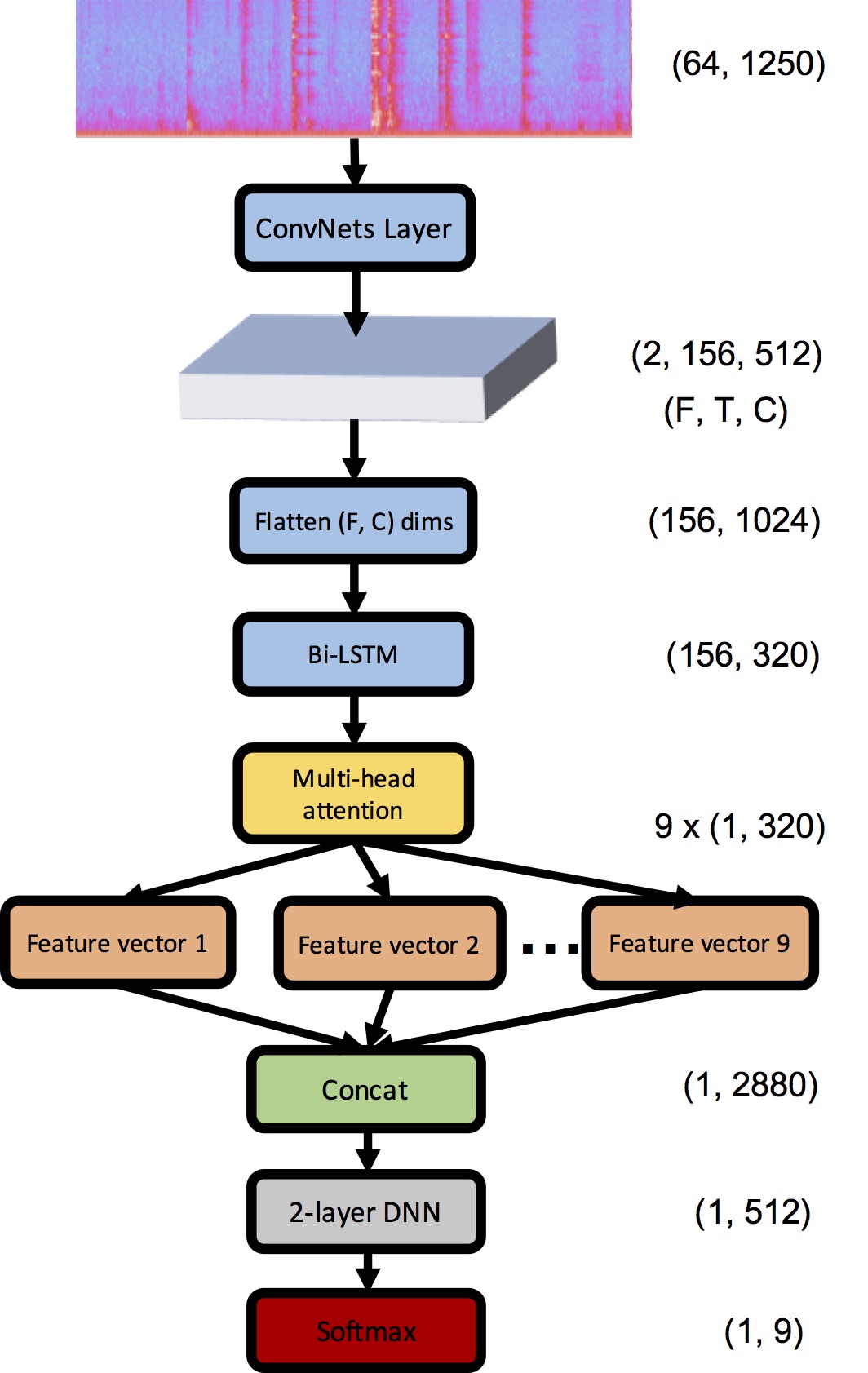}
  \caption{Overall network architecture of our model.}
  \label{f:architecture}
\end{figure}

\subsection{Data augmentation}
\label{s:augmentation}

Proper data augmentation is important for our task, as the class
distribution is very skewed, as shown in Table~\ref{t:class-distribution}:
the smallest class \emph{Vacuum cleaning} contains only $972$ training
samples, compared to  the class \emph{Absence} which contains $18,860$
training samples. Previously, \cite{inoue2018domestic} augmented the data by
randomly selecting two audio clips from the same class, evenly
splitting each clip into five $2$-second segments, and randomly selecting
$5$ out of the $10$ segments, and concatenating them (in random order) to
form a new $10$-second clip as augmented data for training. 

We adopt a similar but simpler approach. We randomly select two clips
within the same class, and from each clip, we cut out a
continuous $5$-second segment, whose starting time is sampled from $[0,
5]$ uniformly at random. The two $5$-second segments are then concatenated
to form a new $10$-second clip for training. Augmentation is done on-the-fly during training, and model
won’t see the same augmented clip two times. We apply this augmentation
strategy to the minority classes, namely \emph{cooking}, \emph{dishwashing}, \emph{eating},
\emph{other}, \emph{social activity} and \emph{vacuum cleaning}.

\subsection{Model architecture}
\label{s:architecture}

For input features, we extract 64D log-Mel
features from the original single-channel audio data, 
with a window size of 16 ms and hop size of 8 ms, followed by
per-utterance mean subtraction. This gives us a 64x1250 feature matrix
per utterance.

Our feature extraction network $f$ starts with VGGish ConvNets
layers~\cite{simonyan2014very}, 
the details of which are given in Table~\ref{t:convnet}. The output of
convolutional layers, with a receptive field of 64 ms in the time axis,
is then fed to a bi-directional LSTM layer~\cite{HochreitSchmid97a} to extract abstract features
with rich temporal information. After that, we apply the multi-head attention
module described in Section~\ref{s:model}. 
This attention module outputs $M$ (e.g., 9) fixed-dimension feature
vectors, which are concatenated to form the final feature vector $\sss$. 
The final classification network $g$ consists of $2$ hidden
layers with $512$ ReLU units~\cite{Nair_10a} each, and a final softmax layer for $9$-way
classification. The overall model architecture is illustrated in Fig.~\ref{f:architecture}.

\subsection{Model training and selection}
\label{s:training}

We adopt the same strategy as the baseline method~\cite{Dekkers2018_DCASE} to define an
``epoch'': we down-sample each class (by random sampling) to
have same number of samples as the smallest class, going through these
samples once constitutes one epoch, and we repeat the down-sampling process before each epoch. 

For training, we used the Adam optimizer~\cite{KingmaBa14a} with minibatches of $200$
utterances and an initial learning rate of $0.001$. Furthermore, we reduce the learning rate by
a factor of $0.5$ every $7$ epochs. We evaluate our model on the dev set every $5$ epochs,
and stop training when the dev set performance, measured by macro F1 score
(which is also the final metric used by the challenge), does not improve
further. 

\begin{figure}[t]
\centering
\begin{tabular}{@{}c@{\hspace{0.02\linewidth}}c@{}}
\rotatebox{90}{\hspace{3em}\large dev macro F1} & 
\includegraphics[width=0.75\linewidth]{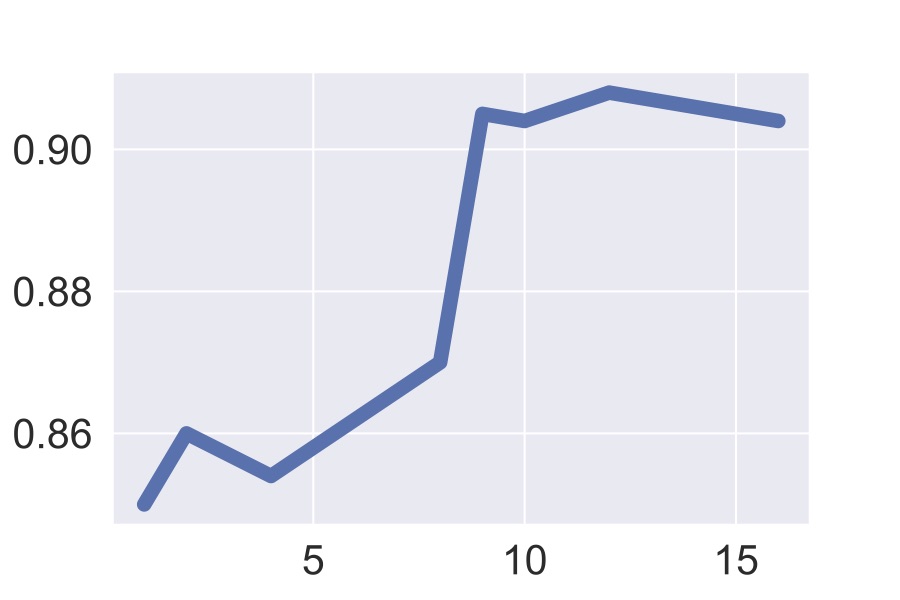} \\
& $M$ (num attn. heads) \\ [1ex]
\rotatebox{90}{\hspace{3em}\large dev macro F1} & 
\includegraphics[width=0.75\linewidth]{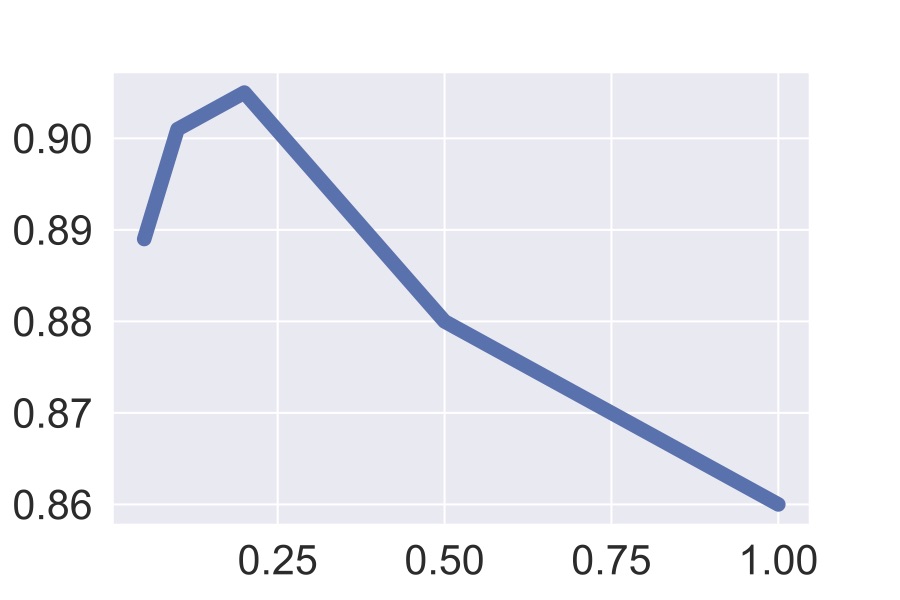} \\
& $\sigma$
\end{tabular}
\vspace*{-1ex}
\caption{Sensitivity analysis of hyperparameters $(M,\sigma)$. Top: dev
 performance for different $M$, with $\sigma=0.2$. Bottom: dev performance
for different $\sigma$, with $M=9$.}
\label{f:sensitivity}
\end{figure}

We tune the two hyperparameters in our method---the number of attention
heads $M$ and the attention shape parameter $\sigma$---by grid search,
based on dev set macro F1 score. 
In Fig.~\ref{f:sensitivity}, we show how the dev
performance changes as we vary one hyperparameter while fixing the other.
From the top plot, we observe that for our task, the performance initially improves as
we increase the number of attention heads $M$, and there is a significant
gain when $M$ reaches $9$ (which coincidentally agrees with the number of
classes), and the performance stabilizes after that. Therefore, we set $M=9$ to balance
performance and computational cost. 
The bottom plot shows the dev performance as we vary $\sigma$, and
$\sigma=0.2$ is chosen for the final model.
Observe there exists a range of hyperparameters for which our model
works similarly well.

\subsection{Results}
\label{s:results}

\begin{table}[t]
\centering
\caption{Final performance (macro F1 score) of our method.}
\label{t:results_table}
\begin{tabular}{@{}|l|r|r|r|r|@{}}
\hline
Class & Baseline & ~\cite{inoue2018domestic} & \caja{c}{c}{Max\\ Pool}  & \caja{c}{c}{Ours\\ \small multi. attn.} \\
\hline
Absence & 0.877 & 0.937 & 0.896 & 0.927 \\
Cooking & 0.930 & 0.915 & 0.935 & 0.938 \\
Dishwashing & 0.772 & 0.865 & 0.829 & 0.866 \\
Eating & 0.812 & 0.870 & 0.849 & 0.880 \\
Other & 0.350 & 0.542 & 0.533  & 0.588 \\
Social activity & 0.966 & 0.979 & 0.977 & 0.979 \\
Vacuum clean. & 0.958 & 0.971 & 0.962 & 0.953 \\
Watching TV & 0.999 & 0.999 & 0.998 & 1.000 \\
Working & 0.814 & 0.887 & 0.822 & 0.884 \\
\hline
Overall & 0.831 & 0.884 & 0.867 & \textbf{0.891} \\
\hline
\end{tabular}
\end{table}

We show our final results on the evaluation set in
Table~\ref{t:results_table}. For comparison, we include also the results
obtained by the baseline model, and the method from the winner of the
challenge~\cite{inoue2018domestic}; the performance of our method (5th column) is on par with the winner's
solution.
For ablation study, we also provide the performance of a variant of our
method which, instead of using multi-head attentions, performs global
max pooling on $f(\X)$ in the time axis, for computing the utterance
representation.
This variant is denoted by ``Max Pool'' in
Table~\ref{t:results_table} (4th column), whose performance significantly
degrades from that of our final model. This demonstrates the
effectiveness of the proposed multi-head attention mechanism.

\section{Attention visualizations}
\label{s:visualizations}

\begin{figure*}[t]
  \centering
  \begin{tabular}{@{}c@{\hspace{0.01\linewidth}}|@{\hspace{0.01\linewidth}}c@{}}
    \includegraphics[width=0.49\linewidth]{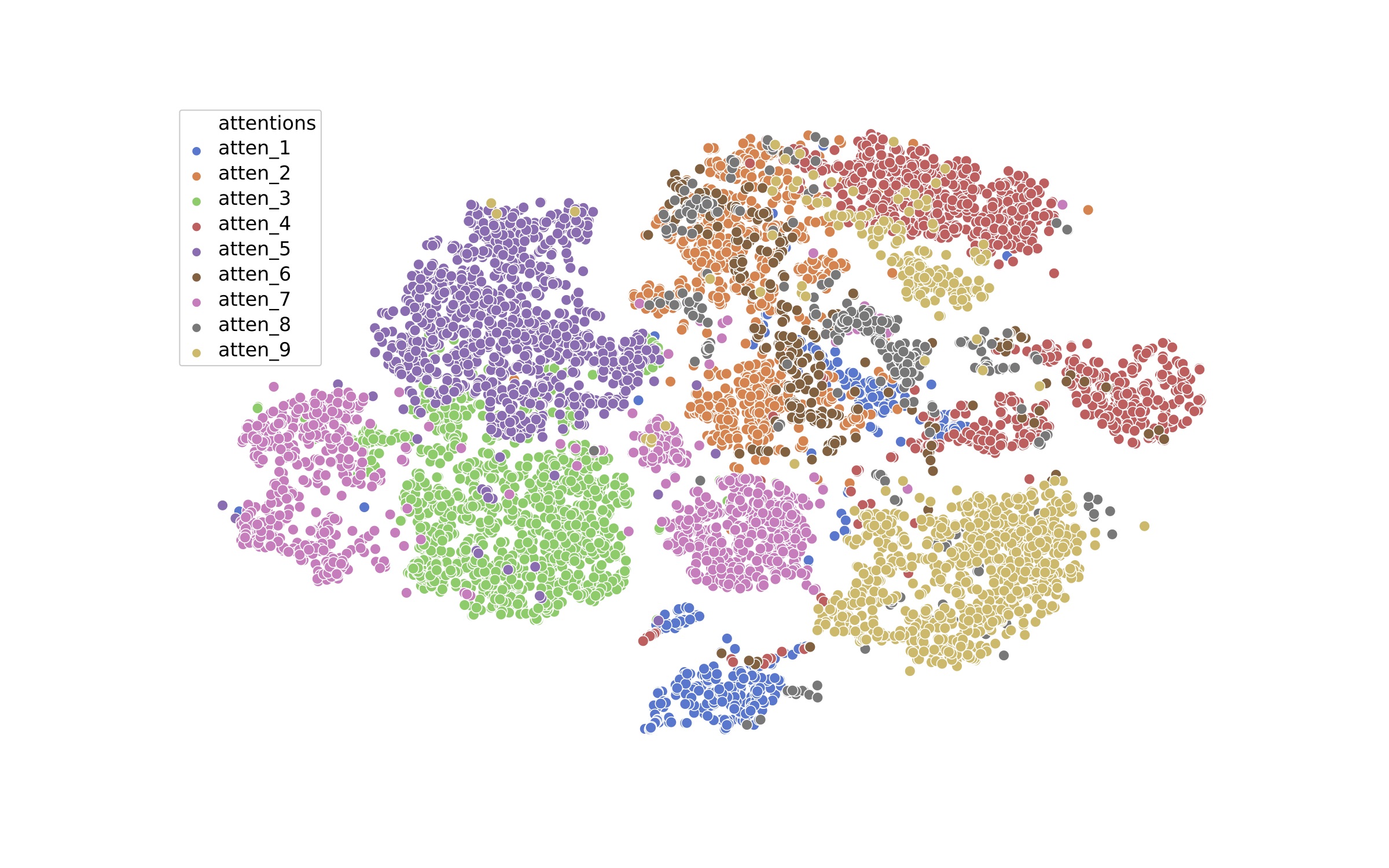} & 
    \includegraphics[width=0.49\linewidth]{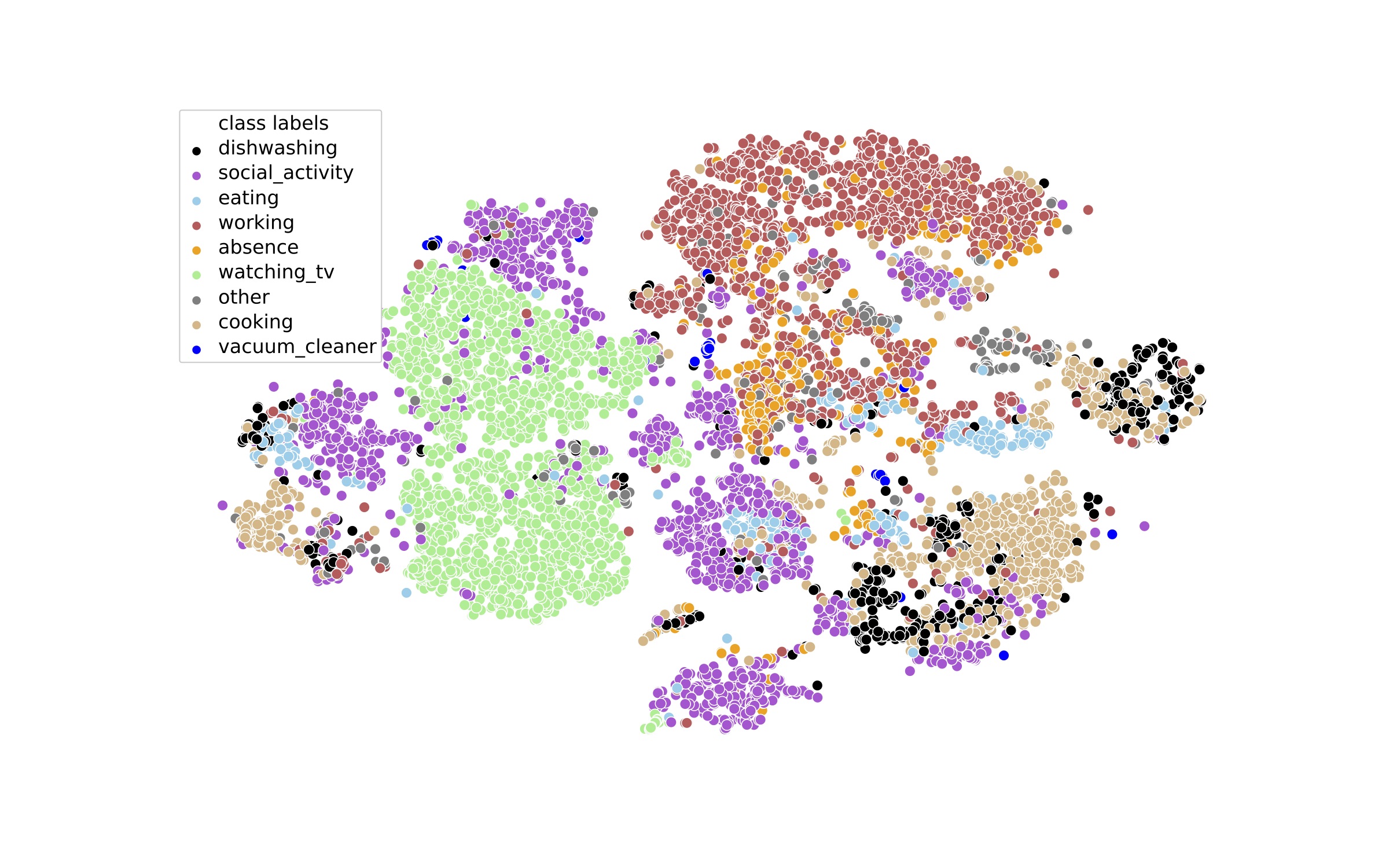}
    \end{tabular}
\vspace*{-1ex}
  \caption{2D t-SNE visualization of $\h_t$'s selected by the attention
    heads. Left plot is colored according to attention head, and right
    plot is colored according to utterance label.}
  \label{f:attention-tsne}
\end{figure*}

\begin{figure*}[t]
  \centering
  \begin{tabular}{@{}c@{\hspace{0.02\linewidth}}c@{}}
    \rotatebox{90} {\hspace*{1.5em} 9 \hspace*{3.5em} 8 \hspace*{3.5em} 7 \hspace*{3.5em} 6 \hspace*{3.5em} 5 \hspace*{3.5em} 4
    \hspace*{3.5em} 3 \hspace*{3.5em} 2 \hspace*{3.5em} 1 }&
    \includegraphics[width=0.90\linewidth,height=0.78\linewidth]{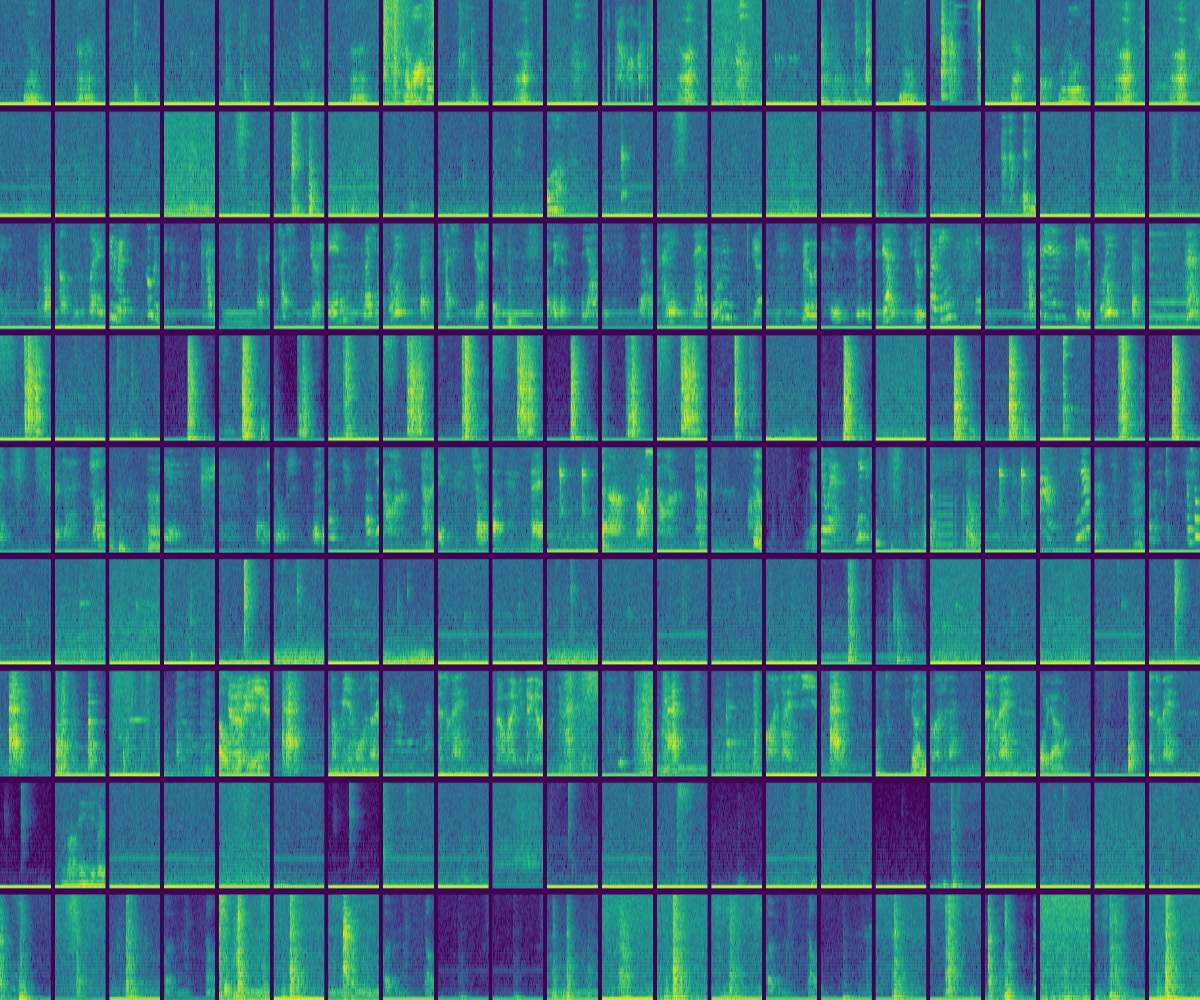}
  \end{tabular}
  \caption{Selection of $1$-second segments selected by the attentions heads.}
  \label{f:attention-samples}
\end{figure*}

In this section, we try to interpret the attention heads. 
Since the attention score $a_t^i$ measures the level of relevance of frame
feature $\h_t$ to event $i$, for each $10$-second audio clip in the dev
set,  we select the $\h_t$ from the bi-directional LSTM outputs that is
most aligned with each event $i$ (i.e., $t = \argmax_{t^\prime}\, a_{t^\prime}^i$), and visualize them with t-SNE~\cite{maaten2008visualizing} in 2D, as shown
in Fig.~\ref{f:attention-tsne}. In the left plot, we color each point according to their
attention head, whereas in the right plot, we visualize each point
according to the class $\h_t$ comes from.

As we can see from the left plot, the $\h_t$'s show strong
clustering associations with each attention heads, implying that the each
attention head focuses more or less on a unique sound pattern (event).
From the right plot, we discover correlations between feature
representations and class labels. 
Notably, the pattern detected by attention 3 almost entirely
belongs to \emph{watching TV}. But in general, the correlation between
attention heads and classes is not one-to-one.
For example, head 5 may also contributes significantly to \emph{watching TV};
on the other hand, some attention heads (such as 4 and 9) cover multiple
closely related classes.

To understand the events each head attends to, for each $\h_t$ selected by
head $i$, we find the corresponding time stamp on the original
$10$-second clip, and select a $1$-second audio segment around the time
stamp and listen to it. A sample of the log-Mel features for these
$1$-second segments are provided in Fig.~\ref{f:attention-samples}. Each row refers to one attention head. 
We find each head attends to one or a few distinctive sound patterns. 
For example, attention heads 3 and 5 mainly detect human speech as well as
media speech and phone conversations; these events are associated with \emph{social activities} and \emph{watching TV} in Fig.~\ref{f:attention-tsne}.
Attention head 4 and 9, on the other hand, detect mostly percussive sounds
like keyboard typing and mouse clicking in \emph{working}, and silverware clanking
and hitting sounds, which are shared by \emph{cooking}, \emph{eating}, and \emph{dishing washing} in Fig.~\ref{f:attention-tsne}.

\section{Conclusion}
\label{s:conclusion}

In this paper, we propose a multi-head attention model which achieves competitive performance for
acoustic scene analysis on DCASE 2018 competition dataset. The multi-head attention
mechanism can discover meaningful representations of distinctive sound
events and locate their appearances in time, given only class labels of the
entire audio clip. Moreover, all parameters in our model can be trained
jointly, in an end-to-end fashion. In future work, we may explore the
proposed model with even more complex scenes and larger number of classes.



\bibliographystyle{IEEEbib}
\bibliography{refs}
\end{document}